\documentclass{llncs}
\input psfig.sty
\usepackage{latexsym}
\usepackage{makeidx}

\newcommand{\EMPTYSET}{\mbox{$\O$}}

\newcommand{\CL}{\mbox{$\varphi$}}
\newcommand{\NCL}{\mbox{$\varphi_{\eta}$}}

\newcommand{\NOT}{\mbox{$-$}}
\newcommand{\NBHD}{\mbox{$\eta$}}
\newcommand{\REG}{\mbox{$\rho$}}

\newcommand{\CALI}{\mbox{${\cal I}$}}

\newcommand{\CALN}{\mbox{${\cal N}$}}

\begin{document}

\bibliographystyle{plain}
\bibstyle{plain}

\title
	{ The Irreducible Spine(s) of Undirected Networks }

\author{
   John L. Pfaltz
	}
\institute{
   Dept. of Computer Science,  University of Virginia 
	}


\maketitle

\begin{abstract}
Using closure and neighborhood concepts, we show that within every undirected network,
or graph, there is a unique irreducible subgraph which we call its ``spine''.
The chordless cycles which comprise this irreducible core effectively characterize the
connectivity structure of the network as a whole.
In particular, it is shown that the center of the network, whether defined by distance
or betweenness centrality, is effectively contained in this spine.

By counting the number of cycles of length
$3 \leq k \leq max\_length$, we can also create a kind of signature that can
be used to identify the network.

Performance is analyzed, and the concepts we develop are
illustrated by means of a relatively
small running sample network of about 400 nodes.
\end{abstract}
\section {Introduction} \label{I}

It is hard to describe the structure of large networks.
If the network has fewer than 100 nodes, then we can hope to draw it as a 
graph and visually comprehend it
\cite{Fre00}.
But, with more than 100 nodes this becomes increasingly difficult.

Simply counting the number $n$ of nodes and number $e$ of edges,
or connections, provides essential basic information.
Other combinatorial measures include counting the number of triangles,
the number of edges incident to a node $v$,
and the total number of nodes such that
$\delta(v) = k$, where
$\delta(v)$ denotes the degree of a node $v$, or
the number of edges incident to $v$.
There exist data representations that effectively keep these kinds of counts,
even in rapidly changing dynamic networks
\cite{LinSouSzw10}.

More sophisticated methods involve treating the defining adjacency matrix
as if it were a linear transformation and employing an eigen analysis
\cite{New06}.
All these techniques convey information about a network.
In this paper, we present a rather different approach.

First in Section \ref{IN} we reduce the network to an irreducible core,
which is shown to be unique (upto isomorphism) for any network.
Then in Section \ref{CKC} we show that the irreducible core is 
comprised exclusively of chordless cycles of length $k$, or $k$-cycles.
The center of the network, whether defined in terms of distance, or 
betweenness centrality
\cite{Bra01},
can always be found in this spine.
In addition, the distribution of these $k$-cycles, $3 \leq k \leq  max\_length$,
can provide a ``signature'' for the network.

Finally, we indicate how spine can be used to estimate other parameters of
the network, such as diameter and number of triangles.
 
\section {Irreducible Networks} \label{IN}

For this paper we regard a network $\CALN$ as an undirected graph on a 
set $N$ of $n$ nodes with a set $E$ of $e$ edges, or connections.
Many of these results can be applied to directed networks as well, but 
we will not explore these possibilities here.
The {\bf neighborhood} of a set $Y$ of nodes are those nodes not in $Y$
with an edge connecting them to at least one node in $Y$.
We denote such a neighborhood by $Y.\NBHD$, that is
$Y.\NBHD = \{ z \not\in Y | \exists y \in Y, (y, z) \in E \}$.
We use this somewhat unusual suffix notation because we regard
$\NBHD$ as a set-valued operator acting on the set $Y$.
By the {\bf region} dominated by $Y$, denoted $Y.\REG$, we mean
$Y.\REG = Y.\NBHD \cup Y$.

In our treatment of network structure, we will make use of the
{\bf neighborhood closure} operator, denoted by $\CL$
\cite{Pfa11}.
For all $Y \subseteq N$, this is defined to be
$Y.\CL = \{ z \in Y.\REG : \{z\}.\REG \subseteq Y.\REG \}$
which is computationally equivalent to
$Y.\CL = Y \cup \{ z \in Y.\NBHD : \{z\}.\NBHD \subseteq Y.\REG \}$.
Readily $Y \subseteq Y.\CL \subseteq Y.\REG$.
Recall that a closure operator $\CL$ is one that satisfies the 3
properties:
(C1) $Y \subseteq Y.\CL$,
(C2) $X \subseteq Y$ implies $X.\CL \subseteq Y.\CL$, and
(C3) $Y.\CL.\CL = Y.\CL$.

Because the structure of large networks can be so difficult to comprehend
it is natural to seek techniques for reducing their size, while still
preserving certain essential properties
\cite{AggWan11,GilLev04},
often by selective sampling
\cite{LesFal06}.
Our approach is some what different.
We view ``structure'' through the lens of neighborhood closure, which
we then use to find the unique irreducible sub-network $\CALI \subseteq \CALN$.

A graph, or network, is said to be {\bf irreducible} if every
singleton subset $\{y\}$ is closed.
A node $z$ is {\bf subsumed} by a node $y$ if 
$\{z\}.\CL \subseteq \{y\}.\CL$.
Since in this case, $z$ contributes very little to our understanding
of the closure structure of $\CALN$, its removal will result in little
loss of information. 
\begin{samepage}
\begin{proposition}\label{p.subsume.1}
Let $y$ subsume $z$ and let $\sigma(x, y)$ denote a shortest path between
$x$ and $y$.
If $z \in \sigma(x, y)$, then there exists $\sigma'(x, y)$ such that
$z \not\in \sigma'$
\end{proposition}
\end{samepage}
\noindent
\begin{proof}
If not, we may assume without loss of generality that $z$ is adjacent to 
$y$ in $\sigma$.
But, then $\sigma(x, z), x \not\in \{y\}.\NBHD$ implies that
$\{z\}.\CL \not\subseteq \{y\}.\CL$.
(Also proven in
\cite{Pfa12}.)
\qed
\end{proof}
In other words,
$z$ can be removed from $\CALN$ with the certainty that
if there was a path from some node $x$ to $y$ through $z$, there will still 
exist a path of equal length from $x$ to $y$ after $z$'s removal.
Such subsumed nodes can be iteratively removed from $\CALN$ without
changing connectivity.
This iterative reduction process we denote by $\omega$.

Operationally, it is easiest to search the neighborhood $\{y\}.\NBHD$
of each node $y$, and test whether $\{z\}.\NBHD \subseteq \{y\}.\REG$
as shown in the code fragment of Figure \ref{CODE}.
\begin{figure}[ht]
{\tt
\begin{small}
\begin{verbatim}
   for_each y in N
      {
      for_each z in y.nbhd
         {
         if (z.nbhd contained_in y.region
            {        // z is subsumed by y
            for_each x in z.nbhd
               remove edge (x, z)
            remove z from network
            }
         }
      }
\end{verbatim}
\end{small}
}
\caption{Key loop in reduction process, $\omega$.
\label{CODE} }
\end{figure}
\noindent
This code is then iterated until there are no more subsumable nodes.
Let $y.\beta$ denote the set of nodes subsumed directly, or indirectly, by $y$.
In a sense these subsumed nodes {\bf belong} to $y$.
Let $\tau(y)$ denote $| y.\beta |$.
Since every node subsumes itself, $\tau(y) \geq 1$.
In our implementation of this code, we also increment $\tau(y)$ by $\tau(z)$
every time node $z$ is subsumed by $y$.
So, $\tau(y) = | y.\beta |$.
Consequently, $\sum_{y \in \CALN.\omega} \tau(y) = n = | \CALN |$.

Before considering the behavior of $\omega$, we want to establish a few
formal properties of the reduced network.
\begin{samepage}
\begin{proposition}\label{p.RED0}
Let $\CALN$ be a finite network and let $\CALI = \CALN.\omega$ be a reduced version,
then $\CALI$ is irreducible.
\end{proposition}
\end{samepage}
\noindent
\begin{proof}
Suppose $\{y\}$ in $\CALI$ is not closed.
Then $\exists z \in \{y\}.\NCL$ implying $z.\REG \subseteq \{y\}.\REG$ or that
$z$ is subsumed by $y$ contradicting termination of the reduction code.
\qed
\end{proof}
\medskip

Two graphs, or networks, $\CALN = (N, E)$ and $\CALN' = (N', E')$ are said to be
{\bf isomorphic}, or $\CALN \cong \CALN'$, if there exists a bijection,
$i:N \rightarrow N'$ such that for all
$x, y \in N$, $(i(x), i(y)) \in E'$ if and only if $(x, y) \in E$.
That is, the mapping $i$ precisely preserves the edge structure, or equivalently
its neighborhood structure. 
Thus,
$i(y) \in i(x).\NBHD'$ if and only if $y \in x.\NBHD$.\footnote
	{
	Note that $i:N \rightarrow N'$ is a normal single-valued
	function on $N$, so we use traditional prefix notation.
	We reserve suffix notation for set-valued operators/functions.
	}

The order in which nodes, or more accurately
the singleton subsets, of $\CALN$ are encountered can alter which points are subsumed and
subsequently deleted.
Nevertheless, we show below that
the reduced graph $\CALI = \CALN.\omega$ will be unique, upto isomorphism.
\begin{samepage}
\begin{proposition}\label{p.RED4}
Let $\CALI = \CALN.\omega$ and $\CALI' = \CALN.\omega'$ be irreducible subsets of a
finite network $\CALN$, then $\CALI \cong \CALI'$.
\end{proposition}
\end{samepage}
\noindent
\begin{proof}
Let $y_0 \in \CALI$, $y_0 \not\in \CALI'$.
Then $y_0$ is subsumed by some point $y_1$ in $\CALI'$ and
$y_1 \not\in \CALI$ else because $y_0.\REG \subseteq y_1.\REG$ implies
$y_0 \in \{y_1\}.\CL$ so $\CALI$ would not be irreducible.
\\
Similarly, since $y_1 \in \CALI'$ and $y_1 \not\in \CALI$,
there exists $y_2 \in \CALI$
such that $y_1$ is subsumed by $y_2$.
Now we have two possible cases; either $y_2 = y_0$, or not.
\\
Suppose $y_2 = y_0$ (which is most often the case), then
$y_0.\REG \subseteq y_1.\REG$ and $y_1.\REG \subseteq y_0.\REG$ or
$y_0.\NBHD = y_1.\NBHD$.
Hence $i(y_0) = y_1$ is part of the desired isometry, $i$.
\\
Now suppose $y_2 \neq y_0$.
There exists $y_3 \neq y_1 \in \CALI'$ such that $y_2.\REG \subseteq y_3.\REG$,
and so forth.
Since $\CALI$ is finite this construction must halt with some $y_n$.
The points $\{ y_0, y_1, y_2, \dots y_n \}$ constitute a complete graph $Y_n$
with $\{y_i\}.\REG = Y_n.\REG$, for $i \in [0, n]$.
In any reduction all $y_i \in Y_n$ reduce to a single point.
All possibilities lead to mutually isomorphic maps.
\qed
\end{proof}
\medskip

\noindent
We call this unique subgraph, the {\bf irreducible spine} of $\CALN$.
In \cite{LinSouSzw10},
Lin, Soulignac and Szwarcfiter,
speak of a ''{\it dismantling} of a graph $G$ as a graph $H$ obtained
by removing one dominated vertex of $G$, until no more dominated vertices remain'';
and similarly conclude that ``all dismantlings of $G$ are isomorphic''.
This is precisely the process we have been describing.

For the remainder of this paper we will use a single example to illustrate 
our approach to describing network structure.
In \cite{New06}
Mark Newman describes a 379 node network in which each node corresponds to
an individual engaged in network research, with an edge between nodes if
the two individuals have co-authored a paper.
\begin{figure}[ht]
\centerline{\psfig{figure=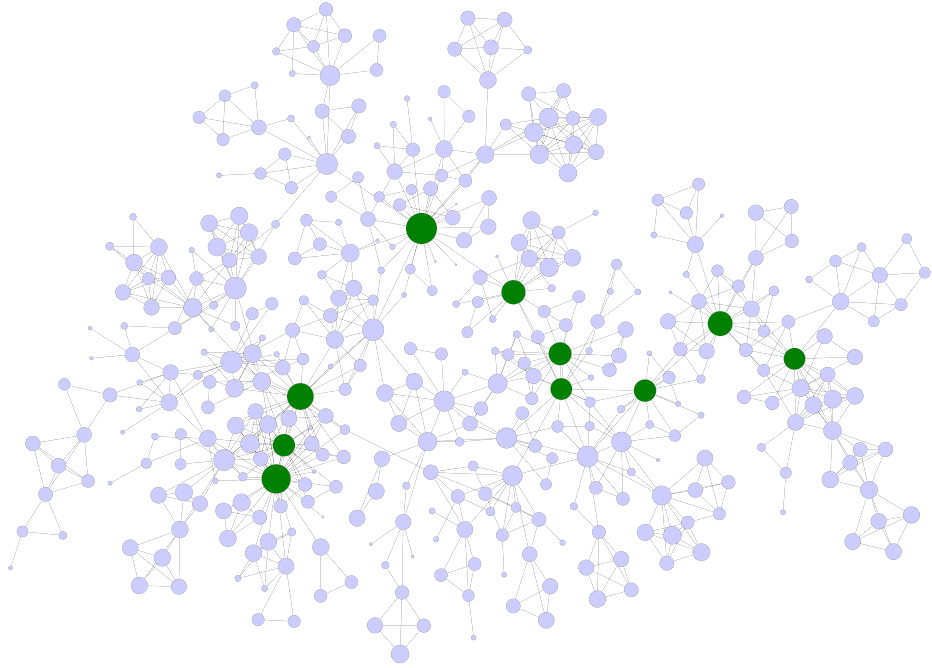,height=3.5in}}
\caption{379 node collaboration network.
\label{NEWMAN.1} }
\end{figure}
The reader is encouraged to 
view an annotated version at
{\tt www.umich.edu/\~\ mejn/centrality}.\footnote
	{
	Similar ``collaboration'' networks can be found in Stamford Large Network
	Database.
	}

\begin{figure}[ht]
\centerline{\psfig{figure=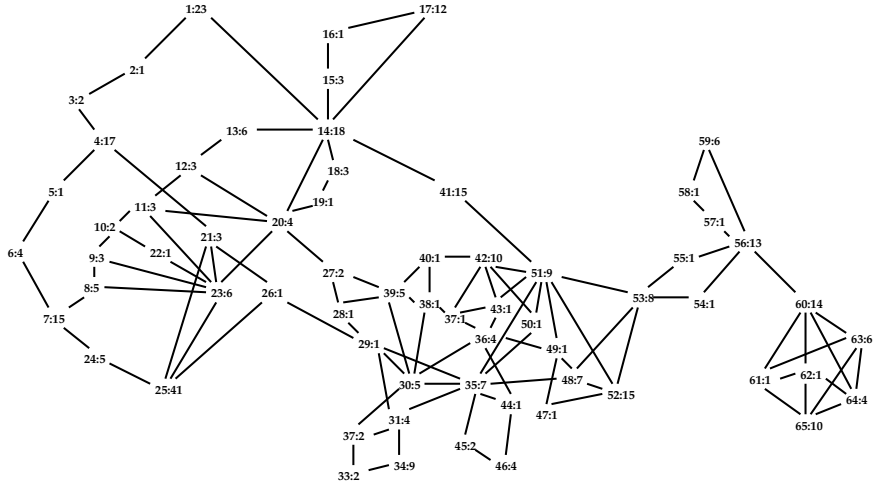,height=2.5in}}
\caption{Reduced 65 node version of Newman's 379 node
co-authorship network.
\label{NEWMAN.3} }
\end{figure}

As described in 
\cite{Pfa12},
we used the code of Figure \ref{CODE} to reduce 
the 379 Newman collaboration network to the 65 node irreducible spine shown in
Figure \ref{NEWMAN.3}.
In our implementation, we keep a count of the number of nodes directly, or
indirectly, subsumed by an irreducible node $y$.
(These counts are displayed in following figures).
In addition, we keep a list of the node identifiers of every subsumed node,
so we can reconstruct a close approximation of the original network.
Thus, this irreducible sub-network $\CALI \subseteq \CALN$ can be regarded as a true
surrogate of $\CALN$ itself.
For simplicity, we have replaced the actual author names with identifying
integers; for example the uppermost node, 1:23, denotes D. Stauffer.
Here 1 is the identifier, $23 = \tau(1)$ denotes the number of individuals in the 
community subsumed by 1.\footnote
	{
	Because there is considerable randomness in the reduction process, 
	several individuals other than Stauffer could have been chosen
	to represent this community.
	Still, the resulting graph would have been isomorphic to Figure \ref{NEWMAN.3}.
	}
By indicating the numbers of individuals/nodes subsumed by a node in 
the reduced version, we suggest the density of the original graph in
this neighborhood.
To further help the reader orient this reduced network with the original,
we observe that 14:18 denotes M. Newman, 23:6 denotes H. Jeong,
25:41 denotes A.-L. Barabasi, 53:8 denotes Y. Moreno and 60:14 denotes J. Kurths.

We must emphasize that we are concerned strictly with the structure of a network,
not its content.
We have chosen this collaboration network solely because it is fairly familiar and
well known.
In no way do we want to suggest that the irreducible sub-network of this section,
or the chordless $k$-cycles described in the following section necessarily contribute
to an interpretation of the significance of the collaboration.
$\CALN$ should be regarded simply as an arbitrary, but relatively complex, network.

What is the computational cost of reducing such a network to its irreducible spine?

The dominant cost is the loop in Figure \ref{CODE} over all $n$ nodes of $N$.
So, it is at least $O(n)$.
Then we have the embedded loop 
\begin{small}
{\tt for\_each z in y.nbhd}.
\end{small}
First, we assume that the degree $\delta(y)$ of each node is bounded
(typically the case in large networks), thus its behavior will still be linear.
In our implementation, all sets are represented by bit strings, with each
bit denoting an element; set operators are thus logical bit operations.
There is no need to loop over the elements of a set.
Consequently, set operations such as union, intersection, or containment
testing, are $O(1)$.
In this case, the entire loop will still be $O(n)$.

\begin{figure}[ht]
\centerline{\psfig{figure=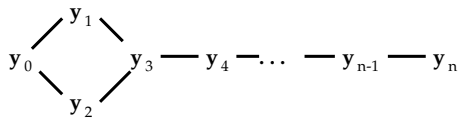,height=0.6in}}
\caption{Reduction, $\omega$, has $O(n^2)$ behavior.
\label{EXAMPLE} }
\end{figure}

However, the loop of Figure \ref{CODE} must be iterated until no more nodes
are subsumed.
It is not hard to create networks in which only one node is subsumed on
each iteration; Figure \ref{EXAMPLE} is a simple example,
if nodes are encountered in subscript order.
So worst case behavior is $O(n^2)$.

The analysis above assumed that the degree of all nodes was bounded.
Suppose not; suppose $\delta(y) \rightarrow n = |N|$.
In this case the node $y$ will subsume many nodes, thereby bounding
the number of necessary iterations.
We have no formal proof for this last assertion, but it appears to 
be true.

Using their H-graph structure, Lin, Soulignac and Szwarcfiter,
show that the cost to dismantle a network is $O(n + \alpha m)$ where 
$\alpha$ denotes the arboricity of $\CALN$
\cite{LinSouSzw10}.
Experimentally,
our reduction, $\omega$, of the Newman collaboration graph to its irreducible
spine shown in Figure \ref{NEWMAN.3} required 5 iterations, with the last over 
the remaining 65 nodes to verify irreducibility.
Other reductions of 4,764, and 5,242, node networks to their 228, and 1,469, node
irreducible spines respectively took 5 and 6 iterations.
In practice, network reduction appears to be nearly linear.
\section {Chordless $k$-Cycles} \label{CKC}

A {\bf cycle} is a closed, simple path
\cite{AgnGre07,Har69}.
A cycle $C = < y_1, y_2, \ldots, y_k, y_1 >$ has length $k$.
For each node $y_i \in C$, $| \{y_i\}.\NBHD | \geq 2$.
The irreducible spine of Figure \ref{NEWMAN.3} has an abundance of
cycles and no nodes $x$ with $| \{x\}.\NBHD | = 1$.


A {\bf chord} in a cycle is an edge/connection $(y_i, y_j) \in E$ where
$j \neq i \pm 1$ (or $i=1, j=k-1$).
A cycle $C$ is {\bf chordless} if it has no chords.\footnote
 	{
 	In \cite{Pfa11,Pfa12}, the author mistakenly used the term ``fundamental
 	cycle'' for the chordless cycles that will be explored in this
 	section.
 	}

\begin{figure}[ht]
\centerline{\psfig{figure=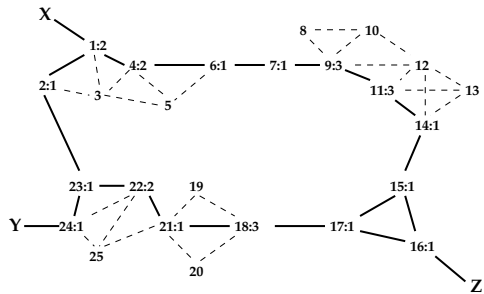,height=1.5in}}
\caption{Granovetter's network with counts of subsumed nodes.
\label{GRANOVETTER} }
\end{figure}

It is the thesis of this paper that these chordless $k$-cycles provide a valuable 
characterization of the structure of a network.
As a small example, consider Figure \ref{GRANOVETTER}
from Granovetter's 1973 article on "weak ties" 
\cite{Gra73},
which has been redrawn
so as to emphasize the chordless 14-cycle.
The nodes $X, Y, Z$ represent other portions of the network.
Readily, describing this subset as a 14-cycle with 9 pendant nodes
is an appropriate characterization.
Our goal with this example is simply to show that
these kinds of chordless cycles arise naturally in the literature and in real life.

\begin{samepage}
\begin{proposition}\label{p.RED1}
Let $\CALN$ be a finite network with $\CALI = \CALN.\omega$ an irreducible version.
If $y \in \CALI$ is not an isolated point then either 
\\
\hspace*{0.1in}
(1) there exists a chordless $k$-cycle $C$, $k \geq 4$  such that $y \in C$, or
\\
\hspace*{0.1in}
(2) there exist chordless $k$-cycles $C_1, C_2$ each of length $\geq 4$ with $x \in C_1$
$z \in C_2$ and $y$ lies on a path from $x$ to $z$.
\end{proposition}
\end{samepage}
\noindent
\begin{proof}
(1)
Let $y_1 \in N_{\CALI}$.
Since $y_1$ is not isolated, let $y_0 \in y_1.\NBHD$, so $(y_0, y_1) \in E$.
With out loss of generality, we may assume $y_0 \in C_1$ a cycle of length 
$\geq 4$.
Since $y_1$ is not subsumed by $y_0$, $\exists y_2 \in y_1.\NBHD, y_2 \not\in y_0.\NBHD$,
and since $y_2$ is not subsumed by $y_1$, $\exists y_3 \in y_2.\NBHD$, $y_3 \not\in y_1.\NBHD$.
Since $y_2 \not\in y_0.\NBHD$, $y_3 \neq y_0$.
\\
Suppose $y_3 \in y_0.\NBHD$, then $< y_0, y_1, y_2, y_3, y_0 >$ constitutes a 
$k$-cycle $k \geq 4$, and we are done.
\\
Suppose $y_3 \not\in y_0.\NBHD$.
We repeat the same path extension.
$y_3.\NBHD \not\subseteq y_2.\NBHD$ implies $\exists y_4 \in y_3.\NBHD$,
$y_4 \not\in y_2.\NBHD$.
If $y_4 \in y_0.\NBHD$ or $y_4 \in y_1.\NBHD$, we have the desired cycle.
If not $\exists\ y_5, \ldots $ and so forth.
Because $\CALN$ is finite, this path extension must terminate with 
$y_k \in y_i.\NBHD$, where $0 \leq i \leq n-3$, $n = |N|$.
Let $x = y_0, z = y_k$.
\\
(2) follows naturally.
\qed
\end{proof}
\medskip

\noindent
The points of those chordal subgraphs still remaining in Figure \ref{GRANOVETTER}
such as the triangle $< 15, 16, 17 >$, are all elements of other chordless
cycles as predicted by Proposition \ref{p.RED1}.
 
\subsection {Centers and Centrality} \label{CC}

A central quest in the analysis of social networks is the identification
of its ``important'' nodes.
In social networks, ``importance'' may be defined with respect to the path structure
\cite{Fre78}.

Let $\sigma (s, t)$ denote a {\bf shortest path} between $s$ and $t$, and let
$d(s, t)$ denote its length, or {\bf distance} between $s$ and $t$. 
Those nodes $C_C = \{y \in \CALN\}$ for which $\delta (y) = \sum_{s \neq y} d(s, y)$ is 
$minimal$ have traditionally been called the {\bf center} of $\CALN$
\cite{Har69}, they are ``closest'' to all other nodes.
It is well known that this subset of nodes must be edge connected.
One may assume that these nodes in the ``center'' of a network
are ``important'' nodes.

Alternatively, one may consider those nodes which ``connect'' many other 
nodes, or clusters of nodes, to be the ``important'' ones. 
Let $\sigma_{st}(y)$ denote the number of shortest paths $\sigma(s, t)$
containing $y$; then those nodes $y$ for which $\sigma_{st}(y)$ is 
$maximal$ are those nodes that are involved in the most connections.
Let $C_B = \{y \in \CALN\}$, for which $\sigma_{st}(y)$ is maximal.
This is sometime called ``betweenness centrality''
\cite{Bra01,Fre78}.
(Note: traditionally, centrality measures are normalized to range between
0 and 1, but we will not need this for this paper.)

In the following sequence we want to show that
nodes with minimal distance and maximal betweenness measures will be found
in the irreducible spine $\CALI$.
This is non-trivial because it need not always be true.
One problem is that, we may have several isomorphic spines,
$\CALI_1, \ldots, \CALI_k$, so we can only assert that $C_C \cap \CALI_j$ and
$C_B \cap \CALI_j$ are non-empty for all $1 \leq j \leq k$.
Second, there exist pathological cases where the centers are disjoint from
$\CALI$.
The network of Figure \ref{EXAMPLE} is an example.
If $n=8$ then $C_C = C_B = y_4$ because
$18 = \delta(y_4) < \delta(y_3) = \delta(y_5) = 19$, and
$24 = \sigma_{st}(y_4) > \sigma_{st}(y_3) = \sigma_{st}(y_5) = 23$.
But, $y_4 \not\in \CALI$.
The conditions of Proposition \ref{p.central.3} will ensure this cannot
happen.
We can assume $\CALI$ is connected, else we are considering one of its 
connected components.

\begin{samepage}
\begin{lemma}\label{l.central.0}
Let $y \in \CALI$ and let $z$ ``belong'' to $y$, $i.e.$ $z \in y.\beta$.
There exists a shortest path sequence $< y_0, \ldots, y_k >$ such that
\\
\hspace*{0.2in}
(a) $y_0 = y$,
\\
\hspace*{0.2in}
(b) $y_k = z$, and
\\
\hspace*{0.2in}
(c) $y_i.\NBHD \subseteq y.\REG = y_i.\NBHD \cup y_i$, $1 \leq i \leq k$.
\end{lemma}
\end{samepage}
\noindent
\begin{proof}
This is a formal property of the subsumption process.
\qed
\end{proof}

\noindent
This sequence need not correspond to the sequence in which nodes are actually
subsumed.

\begin{samepage}
\begin{lemma}\label{l.central.1}
Let $y \in \CALI$, with $z \in y.\beta$ and let $\sigma(s, z)$ be a shortest
path where $s \not\in y.\beta$.
Then there exists a shortest path $\sigma (s,z) = < s, \ldots, y_0, \ldots, y_i, z >$.
\end{lemma}
\end{samepage}
\begin{proof}
Suppose $\sigma (s, z) = < s, \ldots, v, z >$.
Since $z \in y.\beta$, $\exists i, z.\NBHD \subseteq y_i.\NBHD \cup y_i$.
Now $v \in y_i.\NBHD \cup y_i$ hence 
$\sigma (s,z) = <s, \ldots, y_i, z >$ is also a shortest path.
Iterate this construction for $k = i-1,\ldots,0$.
This is also a corollary statement to Proposition \ref{p.subsume.1}.
\qed
\end{proof}

\begin{samepage}
\begin{lemma}\label{l.central.2}
Let $y \in \CALI$ and let $z \in y.\beta, z \not\in y.\NBHD$.
If $s \not\in y.\beta$ then $d(s,z) \geq d(s,y) + 1$.
\end{lemma}
\end{samepage}
\begin{proof}
By Lemma \ref{l.central.1}, $\exists y_k, k \geq 1$ such that 
$z \in y_k.\NBHD$ and $\sigma(s, z) = < s, \ldots, y_0, \ldots, y_i, z >$
is a shortest path. 
Readily $d(s,z) = d(s, y) + i \geq d(s, y) + 1$.
\qed
\end{proof}

\begin{samepage}
\begin{proposition}\label{p.central.1}
Let $y \in \CALI$ with $z \in y.\beta$.
If $z \in y.\NBHD$ then
\\
\hspace*{0.2in}
(a) For all $s, t$, $\sigma_{st}(y) \geq \sigma_{st}(z)$
\\
\hspace*{0.2in}
(b) $\delta (y) \leq \delta (z)$
\end{proposition}
\end{samepage}
\begin{proof}
(a)
Since $z \in y.\NBHD$, and by Lemma \ref{l.central.1}, $i = 0$,
for all shortest paths through $z$, there exists a shortest
path through $y$.
\\
(b) Readily, $z \in y.\NBHD$ and $z.\NBHD \subseteq y.\REG$ implies
$d(s, y) \leq d(s, z)$ for all $s \neq y, z$.
\qed
\end{proof}

\noindent
In this case, $z$ may, or may not, also be in an alternate spine $\CALI'$.
Hence equality is possible in both (a) and (b).

\begin{samepage}
\begin{proposition}\label{p.central.2}
Let $y \in \CALI$ with $z \in y.\beta$.
Let $\sum_{x \in \CALI, x \neq y} \tau(x) \geq \tau(y)$ and let
$\sum_{x\in\CALI, x\in y.\NBHD}\tau(x) \geq \tau(y)$.
If $z \not\in y.\NBHD$ then
\\
\hspace*{0.2in}
(a) For all $s, t$, $\sigma_{st}(y) > \sigma_{st}(z)$
\\
\hspace*{0.2in}
(b) $\delta (y) < \delta(z)$.
\end{proposition}
\end{samepage}
\begin{proof}
(a)
If $s \in y.\beta$ and $t \not\in y.\beta$, then Lemma \ref{l.central.1}
establishes that $\sigma_{st}(y) \geq \sigma_{st}(z)$.
\\
Now suppose that $t \in y.\beta$, then $\sigma(s,t)$ through $z$ need not
imply a shortest path $\sigma(s,t)$ through $y$.
The maximal possible number of such shortest paths occurs when
$y.\beta \NOT y$ is a star graph, such as shown in Figure \ref{STAR}.
\begin{figure}[ht] \label{STAR}
\centerline{\psfig{figure=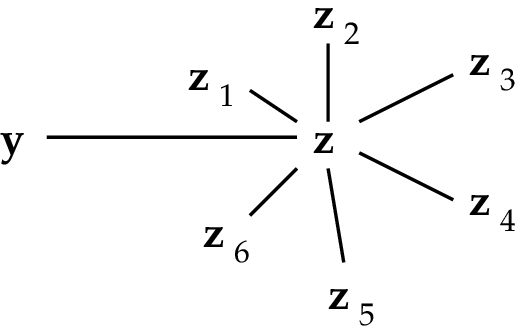,height=0.7in}}
\end{figure}
Let $k = \tau(y) - 2$.
$\exists \ C(k,2) = k\cdot(k-1)/2$ shortest paths $\sigma (s, t)$ through $z$ with
$s,t \neq z$, and $k$ more with $t = z$.
\\
Finally, assume $s, t \not\in y.\beta$.
Let $n = \sum_{x\in\CALI, x\in y.\NBHD}\tau(x) \geq \tau(y)$, a condition of 
this proposition.
This ensures that $\exists \ C(n, 2) + n$ shortest paths through $y$ avoiding $z$.
Since $n > k$, $\sigma_{st}(y) > \sigma_{st}(z)$.
\\
(b)
$\delta(y) = \sum_{s \in y.\beta} d(s, y) + \sum_{t \not\in y.\beta} d(t,y)$
and similarly
$\delta(z) = \sum_{s \in y.\beta} d(s, z) + \sum_{t \not\in y.\beta} d(t,z)$.
Let $k = d(y,z)$, $k \geq 2$.
$\sum_{s \in y.\beta} d(s, y) < \sum_{s \in y.\beta} d(s, z) + k \cdot \tau(y)$.
$\sum_{t \not\in y.\beta} d(t, y) < \sum_{t \in y.\beta} d(t, z) - k \cdot | t \not\in y.\beta|$.
So, provided $\sum_{x \in y.\beta, x \neq y} \tau(x) = | t \not\in y.\beta | > \tau(y)$,
we have
$\delta(y) \leq \delta(z)$.
\qed
\end{proof}

\begin{samepage}
\begin{proposition}\label{p.central.3}
Let $\CALI$ be an irreducible spine of a network $\CALN$ with
centers $C_C$ and $C_B$.
\\
If for all $y \in \CALI$, 
$\sum_{x \in \CALI, x \neq y} \tau(x) \geq \tau(y)$ and 
$\sum_{x\in\CALI, x\in y.\NBHD}\tau(x) \geq \tau(y)$
then
there exist $x_i \in \CALI$ and $y_j \in \CALI$ such that
$x_i \cap C_C$ and $y_j \cap C_B \neq \EMPTYSET$.
\\
Moreover, $C_C \subseteq \cup_i (x_i.\NBHD)$ and
$C_B \subseteq \cup_j (y_j.\NBHD)$.
\end{proposition}
\end{samepage}
\begin{proof}
We compare $y \in \CALI$ with any $z \in y.\beta$.
The first assertion is just a corollary of propositions
\ref{p.central.1}, where $z \in y.\NBHD$, and \ref{p.central.2},
where $z \not\in y.\NBHD$.
\\
The second assertion follows because the inequalities of
Proposition \ref{p.central.2} are all strict.
\qed
\end{proof}

\noindent
The conditions of Proposition \ref{p.central.3} (and Proposition
\ref{p.central.2}) are sufficient to eliminate pathological situations
such as Figure \ref{EXAMPLE};
but are by no means necesary.
In practice, one really only needs that $\CALI$ be sufficiently large,
and that its subsumed sub-graphs not be too unbalanced.

\subsection{Estimation of Other Network Properties} \label{EONP}

The performance of many important network analysis programs is of
order $O(n^k)$, where $k > 1$.
They execute much faster on a small network such as the irreducible 
spine rather than the network itself.
Using $\CALI$ one can often approximate the value with considerable accuracy.
We illustrate by calculating the diameter using Figure \ref{NEWMAN.3}.
Recall that the {\bf diameter} of a network is the maximal shortest
path between any two points.
In \cite{CorLeiRiv96},
the cost to find a diameter using the Floyd-Warshall algorithm is
$O(n^3)$.
This can be reduced to $O(n^2\ log\ n)$ by Johnson's algorithm, but
we know of no better exact solutions.
In Figure \ref{NEWMAN.3} we can do this by hand.\footnote
	{
	This ability is an artifact of this graph structure
	and not generally feasible.
	}

Readily, node 65, in the lower right hand corner is an extreme node.
Expanding out by shortest paths, one finds that node 6 on the left edge is at
distance 13, that is $d(6,65) = 13$, and this is maximal {\it in this 
irreducible spine}.
The center of this subgraph will be nodes at distance 6 or 7 from both 
extremes.
These are nodes 35, 48 and 51, which are necessarily connected in $\CALI$.
Using Proposition \ref{p.central.3} we can assume that at least one of these
is in the actual center of $\CALN$, and that $C_C$ is contained in its
neighborhood.

We continue our estimation of the diameter by considering the subsumed
portions of the network.
What is the nature of the suppressed portions of the network?

Let $y \in \CALI$, $y.\beta$ is a chordal subgraph, where
a subgraph is said
to be {\bf chordal} if it has no chordless cycles of length $\geq 4$.
Chordal graphs are mathematically quite interesting and have been well
studied
\cite{AgnGre07,JacPet90,McK93}.
Succinctly, they can be regarded as tree-like assemblages of complete graphs;
they can be generated by a simple context-free graph-grammar.
In effect, they are pendant tree-like structures that are attached to the
irreducible spine, $\CALI$, at one (or two adjacent) nodes.
Thus $\beta$ is a set-valued operator that associates a pendant
tree of complete graphs with $y$.

Readily, the diameter, $diam_n$ of a chordal graph on $n$ points
satisfies $1 \leq diam_n \leq n-1$, with the lower bound occurring
if $\{y\}.\beta = K_n$, and upper bound when $\{y\}.\beta$ is linear.
In lieu of a better expectation, we will estimate the diameter of 
a pendant chordal graph $\{y\}.\beta$ of $n$ nodes to be $n/2$.
(A much better expectation could be made if both the number of nodes,
and number of edges, were recorded in the reduction process, $\omega$.
This would not be hard.)

With this expected value, we can estimate the length of a maximal
shortest path $(u,v)$ in $\CALN$ $through$ nodes 6 and 65 to be
$d(u, 6) + d(6, 65) + d(65, v)$ or 
$4/2 = 2 + 13 + 5 = 10/2$, or $d(u, v) = 20$, where $u \in \{6\}.\beta$
and $v \in \{65\}.\beta$.

However, this $(u, v)$ path does not appear to actually be the longest path
($i.e.$ diameter).
For the adjacent node 7, $\{7\}.\beta = 15$.
So for $u \in \{7\}.\beta$ we estimate $d(u, v)$ to be $7.5 + 12 + 5 = 24.5$.
And for node 25, $\{25\}.\beta = 41$, so for $u \in \{25\}.\beta$
we would have $diam(\CALN) \approx d(u, v) = 20.5 + 10 + 5 = 35.5$,
which seems to be maximal.
It would be interesting to know what the actual diameter of the original
379 node collaboration graph is.

A similar process can be used to count triangles in the network
\cite{TsoDriMic11}.

\subsection {Network Signatures} \label{NS}

If we count the cycles in the reduced Newman collaboration graph of
Figure \ref{NEWMAN.3}, we get the following enumeration.
\begin{figure}[ht]
\centerline{\psfig{figure=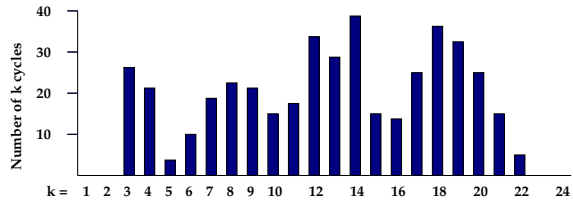,height=1.0in}}
\caption{Distribution of $k$-cycles in Figure \ref{NEWMAN.3}.
\label{BAR.GRAPH} }
\end{figure}
This distribution of chordless cycle lengths may serve as a kind
of spectral analysis, or ``signature'' of the network.
Much more research is needed to determine the value of these
signatures for discriminating between networks.
For example, at SocInfo 2012 in Lausanne, Switzerland, it was suggested
that $CC = \sum_k k \times n_k / |n|$, where $n_k$ is the number
of $k$ cycles, might serve as a measure of connective complexity.

\begin{figure*}[ht]
\centerline{\psfig{figure=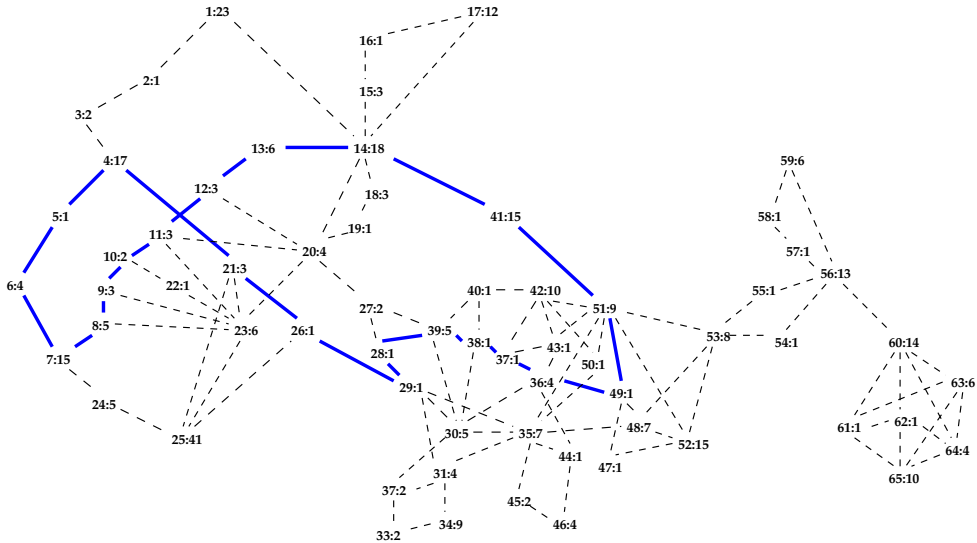,height=2.8in}}
\caption{A maximal chordless cycle in the reduced Newman graph
of Figure \ref{NEWMAN.3}.
\label{NEWMAN.4} }
\end{figure*}

As we see, there are still 26 triangles in this reduction; such
graphs are not ``triangle-free''.
However, it is the 5 chordless cycles of maximum length that are of most interest.
We might call them ``major cycles''.
One of them is:
$<$ 4, 5, 6, 7, 8, 9, 10, 11, 12, 13, 14, 41, 51, 49, 36, 37,
38, 39, 28, 29, 26, 21, 4 $>$.
This one has been emboldened in Figure \ref{NEWMAN.4}
where we emphasize this 22-cycle of maximal
length, while suppressing other aspects of this network.

The reduction process, $\omega$, retains only nodes on, or between, chordless 
$k$-cycles, $k \geq 4$ (Proposition \ref{p.RED1}).
It eliminates the ``chordal'' subgraphs of $\CALN$.
It can be argued that by removing the chordal portions of a network $\CALN$,
$\omega$ is only deleting well understood sub-sections that can be reasonably
well simulated and ``re-attached'' to the irreducible spine.
Just retaining the size of these subsumed subgraphs permits calculation of 
certain global attributes, such as diameter and centrality, as described
earlier.

Looking at Figure \ref{NEWMAN.4} we see a similar process taking place.
The entire subgraph consisting of nodes $\{ 54, 55, \ldots, 65 \}$ is another
pendant portion which will be ignored if one concentrates solely
on the longest $k$-cycles.
The edge/connection $(56, 60) \in E$ is retained, solely because it
connects the two 4-cycles among nodes $\{60, \dots, 65\}$ to the main body,
as described in Proposition \ref{p.RED1}.

\begin{figure*}[ht]
\centerline{\psfig{figure=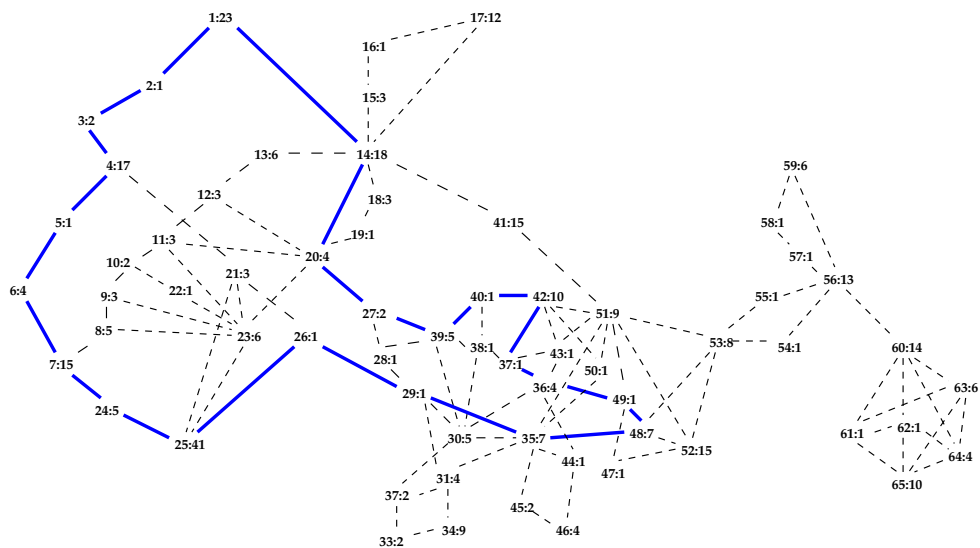,height=2.8in}}
\caption{Another maximal chordless cycle in the reduced Newman graph
of Figure \ref{NEWMAN.3}.
\label{NEWMAN.5} }
\end{figure*}

%
The 22-cycle shown in Figure \ref{NEWMAN.4} is only one of five longest
chordless cycles; a second is shown in Figure \ref{NEWMAN.5}.
As can be seen, it involves other paths.

If the five longest $k$-cycles are intersected, we discover that
10 nodes occur in all.
They are $\{ 4,$ 5, 6, 7, 14, 26, 29, 36, 39, 49 $\}$.
And, four connections appear in all longest cycles,
they are $\{$ (4, 5), (5, 6), (6, 7), (26, 29) $\}$.
The implications of this requires further study.
\section {Summary}

One should have many tools on hand to understand the nature of large graphs,
or networks.
In this paper we have presented one that is rather unusual, yet also rather
powerful.
Even so, it must be observed that 
the reduction, $\omega$, of graphs will always be of mixed value.
Some graphs, for example chordal graphs, will reduce to a single 
node.
This in itself conveys considerable information, but in this case other 
kinds of analysis are clearly more appropriate.
Nevertheless, for many of the kinds of networks one encounters in
real situations, reducing the network to its irreducible spine is a 
quick, easy first step.

Because the irreducible spine, $\CALI$, is effectively unique, further analysis
of it is a valid way of getting information about the original
network.
It is a ``reliable'' surrogate.
It preserves connectivity and path centrality concepts.
Consequently, this kind of analysis with respect to closed sets
can provide valuable insights into the 
nature, and the structure, of the network.

We believe that by counting edges as well as nodes in the subsumed
chordal portions we can get much tighter bounds on the diameters
and triangle counts in these subgraphs, and thus in the entire
network.
This, and further exploration of the idea of connective complexity
are some of future research projects arising from this work.

{\small
\bibliography{/home/jlp/MATH.d/BIB.d/mathDB,/home/jlp/MATH.d/BIB.d/pfaltzDB,/home/jlp/MATH.d/BIB.d/otherDB,/home/jlp/MATH.d/BIB.d/orlandicDB,/home/jlp/MATH.d/BIB.d/haddletonDB,/home/jlp/MATH.d/BIB.d/databaseDB,/home/jlp/MATH.d/BIB.d/evsciDB,/home/jlp/MATH.d/BIB.d/aiDB,/home/jlp/MATH.d/BIB.d/closureDB,/home/jlp/MATH.d/BIB.d/dataminingDB,/home/jlp/MATH.d/BIB.d/conceptDB,/home/jlp/MATH.d/BIB.d/accessDB,/home/jlp/MATH.d/BIB.d/logicDB,/home/jlp/MATH.d/BIB.d/closureappDB,/home/jlp/MATH.d/BIB.d/gtransDB,/home/jlp/MATH.d/BIB.d/topologyDB,/home/jlp/MATH.d/BIB.d/socialDB}
}

\end{document}